\documentclass[1p]{elsarticle}

\usepackage{hyperref}
\usepackage{subfig}
\usepackage{amsmath}
\usepackage{booktabs}
\usepackage{graphicx}
\usepackage{multirow}

\usepackage[colorinlistoftodos,prependcaption,textsize=tiny]{todonotes}
\usepackage{blindtext}
\presetkeys%
    {todonotes}%
    {inline,backgroundcolor=yellow}{}

\journal{Simulation Modelling Practice and Theory}

\begin{document}

\begin{frontmatter}

\title{The Simulation Model Partitioning Problem: an Adaptive Solution Based on Self-Clustering (Extended Version)\footnotemark[0]}

\footnotetext[0]{The publisher version of this paper is available at \url{http://dx.doi.org/10.1016/j.simpat.2016.10.001}.
\textbf{{\color{red}Please cite this paper as: ``Gabriele D'Angelo. The Simulation Model Partitioning Problem: an Adaptive Solution Based on Self-Clustering. Simulation Modelling Practice and Theory, Elsevier, Volume 70, January 2017''.}}}

\author{Gabriele D'Angelo\corref{cor1}}
\ead{g.dangelo@unibo.it}

\address{Department of Computer Science and Engineering. University of Bologna, Italy.}

\cortext[cor1]{Corresponding Author. Address: Department of Computer Science and Engineering. University of Bologna. Mura Anteo Zamboni 7. I-40127, Bologna. Italy. Phone +39 051 2094511, Fax +39 051 2094510} 

\begin{abstract}
This paper is about partitioning in parallel and distributed simulation. That means decomposing the simulation model into a number
of components and to properly allocate them on the execution units. An adaptive solution based on self-clustering, that considers both communication reduction and computational load-balancing, is proposed. The implementation of the proposed mechanism is tested using a simulation model that is challenging both in terms of structure and dynamicity. Various configurations of the simulation model and the execution environment have been considered. The obtained performance results are analyzed using a reference cost model. The results demonstrate that the proposed approach is promising and that it can reduce the simulation execution time in both parallel and distributed architectures.
\end{abstract}


\end{frontmatter}

\section{Introduction}\label{sec:introduction}

Simulation is a widely used technique for the performance evaluation of complex systems. The model resulting from the system abstraction is often so complex that an approach based on a single execution unit is not feasible. Thus, the usage of multiple interconnected execution units is preferred, mostly for speed issues. The partitioning problem is about decomposing the simulation model into a number of components and properly allocating them on the execution units. This allocation procedure has at least two goals: the computation load in the execution architecture has to be balanced as much as possible while the communication overhead among the components has to be minimized~\cite{bagrodia98}. This problem is very significant in Parallel and Distributed Simulation (PADS)~\cite{fujimoto99} but can be generalized to all distributed architectures made by a set of interacting components.

In the PADS research area, a lot of work has been done on Data Distribution Management (DDM)~\cite{boukerche02}, that is the mechanism for distributing state updates and interaction information in a distributed simulation. Even if there are some similarities and correlations between partitioning and DDM, they pertain to different problems in specific logical layers. More in detail, the DDM is about filtering all the interactions that are produced in the simulated model. In other words, it is a mechanism for the efficient matching of the production of information and the expressions of interest. In this case, the goal is to reduce the communication overhead, delivering the interactions only to the parts of the execution architecture that are really interested in them. On the other side, partitioning is about the allocation of the simulation parts on the parallel/distributed execution architecture. Among others, efficient DDM mechanism, synchronization protocols and partitioning are key requirements to obtain an effective PADS.

The partitioning problem is made complex by some factors that have to be taken in account. Firstly, it is not realistic to assume that all the simulation model components are homogeneous and will maintain exactly the same behavior for the whole duration of the simulation. In many simulations, the model components are very heterogeneous both in terms of computation and communication requirements. Secondly, turning our attention to the execution architecture, it is quite rare (and costly) to build up execution architectures that are completely homogeneous in terms of hardware and performance. It is much more cost effective to have execution clusters made of a mixture of commercial off-the-shelf hardware. Even in case of homogeneous execution clusters, the presence of any background computation and communication affects the performance of each node in the cluster and therefore should be taken in account by the model partitioning. In other words, every static partitioning would be unable to satisfy all such requirements and, in most cases, would lead to unsatisfactory performance~\cite{gda-simpat-2014}.\\

In this paper, we propose an adaptive partitioning approach that is based on the dynamic re-allocation of simulation model components. A simple assumption is at the base of our idea: the components of a system interact following a communication pattern that is not random. In fact, in most real world systems, there are physical and functional characteristics that have a clear effect on the interaction dynamics. We think that it is possible to exploit such physical and functional characteristics to rearrange the partitioning of simulation components obtaining significant advantages. In practice, we propose an adaptive approach that is based on the re-allocation (i.e. migration) of simulation components. Under the implementation viewpoint, a PADS middleware provides the usual services to the simulation model and, moreover, supports the transparent migration of simulation components (in the form of agents) among the execution architecture. The approach that we propose is based on the self-clustering of autonomous agents. More in detail, the migrations are triggered by specific heuristics that are evaluated in each agent and based only on local data. If we consider speaking of the communication aspects alone, it is clear that the adaptive partitioning mechanism is fruitful only if the communication cost reduction obtained through the re-allocation is higher than the related migration cost. 
In this work we aim to evaluate the feasibility of this approach and to find what are the system characteristics that can lead to a performance gain.\\

The wireless networks are a good example: due to the physical characteristics of the electromagnetic spectrum, the interaction among wireless devices is usually location dependent. In other words, all the devices that are near the transmitter will receive its packets. More far away, only interferences will be received. And finally, very far devices will not be affected at all. In mobile wireless networks, at every moment the mobility of nodes changes the neighbors of each node and therefore the network topology. Under a simulation point of view, this means that the communication pattern among the simulation components always changes during the whole simulation lifespan. 

The remainder of this paper is organized as follows: Section~\ref{sec:background} reviews the main principles at the basis of our work such as: discrete event simulation, parallel/distributed simulation and the adaptive self-clustering mechanism. Section~\ref{sec:costanalysis} introduces a cost analysis for the performance evaluation of both static and adaptive parallel/distributed simulations. The adaptive migration mechanism, that we propose, is described in detail in Section~\ref{sec:adaptive}. In Section~\ref{Performance}, the performance of the self-clustering mechanism is investigated. Section~\ref{sec:relatedwork} discusses the related work and, finally, in Section~\ref{sec:conclusions} we provide some final remarks.

\section{Background}\label{sec:background}

Many simulation paradigms have been proposed, each one with some benefits and drawbacks. Among them, Discrete Event Simulation (DES)~\cite{Law:1999:SMA:554952} gained attention for being powerful in terms of expressiveness and easy to use. In a DES, the advance of the modeled system is given by a chronological sequence of events. Each event is a change in the system state and happens at an instant in time. Hence, the evolution of the simulation is given by the creation, delivery and computation of events. 
In its simplest form, a DES is implemented by some state variables (i.e.~to describe the modeled system), an event list (i.e.~the pending events that have to be processed), and a global clock (i.e.~the current simulation time)~\cite{Law:1999:SMA:554952}.

In a sequential (i.e. monolithic) simulator, a single Physical Execution Unit (PEU)\footnote{For better readability, the main symbols and acronyms used in the paper are reported in Table~\ref{tab:symbols}.} is responsible for all the state variables of the simulation model and the processing of all events. This doesn't happen in a PADS, in which the simulation model is distributed among a set of interconnected PEUs that can be multi-core CPUs, shared memory multiprocessors, LAN-based clusters or, more recently, cloud infrastructures~\cite{gda-simpat-2014}. In a PADS the delivery of events is obtained through the exchange of messages. It is clear that an execution architecture based on a set of interconnected PEUs has some characteristics that have to be considered in the model partitioning. Firstly, the PEU can be heterogeneous in terms of execution speed, and different interconnection technologies offer very different levels of performance (i.e.~bandwidth, latency and jitter). In practice, this means that the communication among the model components in different parts of the execution architecture can be more or less costly. For example, two execution cores in the same CPU have bandwidth and latency that is way better than two hosts in the same LAN. Secondly, the performance of the execution architecture can vary a lot during the simulation execution due to background load, network congestion and so on. 

In the PADS approach, each PEU manages a part of the simulation model (i.e.~a simulation component) that is implemented by a Logical Process (LP)~\cite{fujimoto99}. In our vision, each LP acts as the container of some Simulated Entities (SEs). That is, the simulation model is partitioned in its basic components (i.e. the SEs) that are properly allocated among the LPs. The behavior of the simulated system is modeled by interactions (i.e.~events) among the SEs. It is pretty obvious that defining the appropriate granularity of the SEs is hard. Using too small SEs increases the simulator management overhead and can boost the communication overhead. Conversely, having a few big sized entities means less control in many aspects such as load balancing. Very often, the SEs granularity is induced by the characteristics of the simulated system. 

\begin{figure}[ht]
\centering
\includegraphics[width=9.0cm]{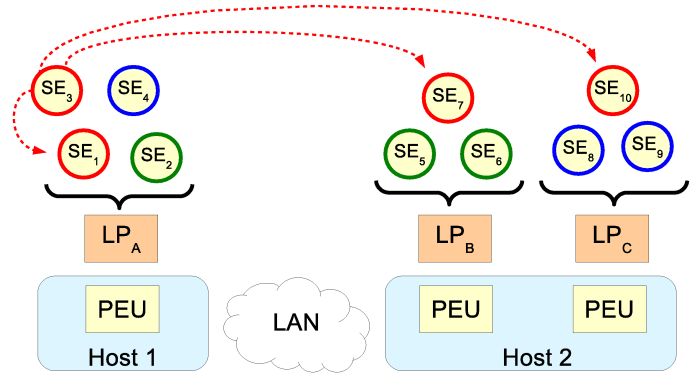}
\caption{Layered structure of a distributed simulator.}
\label{fig_structure}
\end{figure}

\begin{figure}[ht]
\centering
\includegraphics[width=9.0cm]{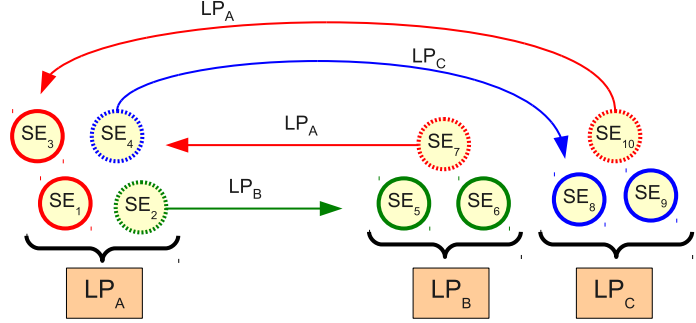}
\caption{Migrations triggered for enhancing the model partitioning.}
\label{fig_migration}
\end{figure}

Figure~\ref{fig_structure} shows an example in which two hosts implementing a distributed simulation are interconnected by a LAN. $Host 1$ has a single PEU (e.g.~a single core CPU) and allocates one LP, $Host 2$ has $2$ PEUs (e.g.~a dual core CPU) and each of them manages a LP. Every LP acts as the container of a set of SEs that interact using messages (depicted as dotted lines in the figure). The colors used to draw the SEs represent the different interaction groups (that is the groups of SEs that, for some reason, interact with high frequency). For example, in the simulation of a MANET we can expect to have many interaction groups that are based on the proximity of wireless devices. The broadcasting of a new message from SE$_3$ to its whole interaction group (i.e.~the reds in Figure~\ref{fig_structure}), in practice requires a single communication inside the LP$_A$ (and its PEU) and two LAN-based communications (to reach SE$_7$ and SE$_{10}$ respectively). If the interaction groups are supposed to last for some time, a 
much better solution would be to cluster each 
group in the same LP. For example, a better partitioning can be obtained with the reallocation (i.e.~migration) of some SEs, as depicted in Figure~\ref{fig_migration}. In this very simple example, migrating 4 SEs leads to a very good clustering. In fact, the new allocation reduces the amount of necessary LAN-based communications, in other words a better communication clustering is implemented. Obviously, in most cases the interaction groups are not static since the simulation model evolution will require the interaction of different SEs (e.g. due to the wireless devices mobility). This means that, in order to enhance the partitioning, further adjustments will have to be done during the whole course of the simulation. In other terms, our goal is to find the logical partitions formed in the simulated model and to reflect them in a physical partitioning (in the execution architecture).

As said before, in our vision the partitioning problem is both about the communication clustering and about the computational load balancing. Therefore, we think that all the proposed solutions to the problem must address both aspects and this is our approach to the problem. However, in this paper we intend to focus on the communication aspects since our goal is to determine if and when the communication clustering can lead to a performance gain, considering also that the computational load balancing would complicate the analysis and reduce readability. In Section~\ref{subsec:loadbalancing}, it will be briefly explained how the proposed approach has been extended to deal with computation issues.

\section{Cost Analysis}\label{sec:costanalysis}

In the previous sections, the partitioning problem has been slightly redefined such that now it is about ``what'' SEs have to be allocated ``where'' in the execution architecture (that is, in which LP). Before going on with the description of the proposed method we need to discuss more in detail the cost of a simulation run.

Let's start with the simplest case, that is the sequential (monolithic) simulation: as said before the whole simulation model is managed by one PEU (e.g. a single core CPU). The Total Execution Cost ($TEC$) can be defined as the amount of time that is needed to complete a simulation run. In this case, the $TEC$ is composed of the Model Computation Cost ($MCC$) and the Local Communication Cost ($LCC$).

\begin{equation}
\label{TotalExecutionCost}
	TEC = MCC + LCC
\end{equation}

In a discrete event simulation, this means that all the time is spent delivering messages among the simulated entities and computing them (e.g.~updating the state variables). The next step is to define what happens in a PADS, where the simulation run is executed by a set of $N$ coordinated PEUs, each one running a single LP\footnote{In our approach, running more LPs on the same PEU has the effect to increase the overhead (due to context switches of LP processes). If a processor is made by multiple execution resources each of them should be considered a PEU and should accommodate a LP.}.

\begin{equation}
\label{TotalExecutionCost-pads}
	TEC = \frac{MCC}{f(N)} + CC
\end{equation}

Where $f(N)$ is a function such that $f(N)>N$, that means that there is a sequential fraction of the simulator that can not be parallelized. Now, given the distributed nature of the execution architecture we have a generic Communication Cost ($CC$) that needs to be described more in detail. In fact, $CC$ is composed of Synchronization Cost ($SC$), Model Interaction Cost ($MIC$) and some Middleware Management Cost ($MMC$).

\begin{equation}
\label{TotalExecutionCost-pads-exploded}
	TEC = \frac{MCC}{f(N)} + (SC + MIC + MMC)
\end{equation}

This means that, to obtain meaningful results, the LPs have to be properly synchronized ($SC$) and furthermore some overhead is due to the management of the software middleware ($MMC$) used to implement the PADS. The MIC is about the cost that is paid for delivering the interactions among the SEs composing the model. Each interaction has a cost that depends on many factors, for example the size of the message used for delivering the interaction and the destination of the message. This last point is crucial, it makes a big difference if the interaction is delivered to the same LP or not. In practice, $MIC$ is composed of two different terms: local and remote communications. In this case, local refers to the LP that sends the interaction. That is, if the destination SE is in the same LP of the sender, then it is possible to say that such a communication is local, in all other cases it is referred as remote.

\begin{equation}
\label{ModelInteractionCost}
	MIC = LCC + RCC
\end{equation}

The ratio between the Local Communication Cost ($LCC$) and the Remote Communication Cost ($RCC$) has a strong impact on performances given that the cost of remote interactions is orders of magnitude higher than the local ones. The intra-LP interactions (i.e. local) are implemented using random-access memory with low latency and high bandwidth. In the case of inter-LP interactions, the network performance depends on the used interconnection technology and its current load. For example, a set of LPs allocated on multi-core/multi-processors can be able to access some shared memory, a LAN-based cluster needs to rely on some network technology (e.g.~Gigabit Ethernet) and, in some in cases, Internet is the only option. It is possible to further specify the communication cost assuming different classes. For example, the communication among LPs on the same CPU would fall in the low cost class, the LAN-based communication in an intermediate class and the Internet-based communication would be in a high cost class. Even if this can be useful for implementation purposes, our analysis is made simpler assuming only two classes: local and remote communications.

Given the cost difference between these two communication types, the best performance can be obtained maximizing the local interactions. For this reason, we propose a migration mechanism that re-allocates the SEs among the LP. In other words, a mechanism that changes the partitioning configuration. This aims to cluster the SEs that interact frequently in the same LP, reducing the use of costly inter-LP communications. In Equation~\ref{TotalExecutionCostWithMigration} all these aspects are considered.

\begin{equation}
\label{TotalExecutionCostWithMigration}
	TEC = \frac{MCC}{f(N)} + (SC + LCC + RCC + MMC) + MigC
\end{equation}

The new term ($MigC$) introduced in the previous equation is the Migration Cost. It means that the total execution cost of the PADS now includes the computation and communication costs paid for the reallocation of SEs. More in detail, $MigC$ can be seen as composed of few addends:

\begin{equation}
\label{MigrationCost}
	MigC = MigCPU + MigComm + Heu
\end{equation}

In which, $MigCPU$ is the generic computation time used to implement the migration (e.g. the serialization of data structures of migrating SEs). $MigComm$ is the cost of the transfer of the migration messages (i.e. the messages used to migrate the SEs and their internal state between LPs). Finally, the $Heu$ is the cost of the heuristic function used to evaluate the simulator execution at runtime and to trigger re-allocations. It is worth noting that each LP has only a local view of both the simulation model and the execution architecture. This means that all the reallocation decisions have to be taken considering only partial information and without a full knowledge of the system. As an alternative, a single point of centralization could be added or all the useful data could be broadcasted to all LPs. In our opinion, both these ways are very antithetic to the PADS approach and with many scalability issues.

Many different approaches can be followed to enhance the partitioning in adaptive PADS. Due to the complexity of the problem to solve (e.g.~the number of SEs, LPs and timesteps), it not possible to find the best allocation of SEs to LPs using an analytical approach. Moreover, the knowledge about the interactions of each SE during the simulation is not available a priori. For this reason, we have chosen to use an adaptive mechanism that is based on simple self-clustering heuristics. In the following of this paper, the proposed mechanism will be introduced and investigated using the cost analysis described in this section as reference. In future works, it will be possible to compare different solutions using this cost model.

\section{Adaptive Migration Mechanism}\label{sec:adaptive}

In the previous section, the main costs of a PADS have been defined at high level. The adaptive migration mechanism leads to some modifications in Equations~\ref{TotalExecutionCostWithMigration} and~\ref{MigrationCost}. The goal is to pay some extra communication and computation costs (i.e.~the migrations) with the aim to reduce the rate of costly inter-LP communications while opting for cheaper intra-LP communications.

There are a couple of issues to consider. First of all, we will see how to design and implement such kind of mechanism. Secondly, its outcomes will be evaluated in different situations and configurations. If the cost of migrations is less than the savings obtained by clustering, then there is a speedup. The problem is that the balance between these two factors depends on many parameters that are both in the execution architecture (e.g. the cost of communication among PEUs) and in the simulated model (e.g. the state size of the SEs and their interaction pattern). In the following of this section, the proposed approach is described in detail. The evaluation is left to the next section.

\subsection{GAIA basics}

The basic idea is that, in most systems, the interactions among its parts are not randomly distributed and that some communication patterns can be observed. These patterns derive from the nature of the systems to be simulated and are reflected on the corresponding simulation models. When this assumption is true, it is possible to analyze such communication patterns and to find the simulation components that are involved in the communication. Each time a group of interacting components is observed, it is evaluated the potential performance gain resulting from clustering.

\begin{figure}[ht]
\centering
\includegraphics[width=9.0cm]{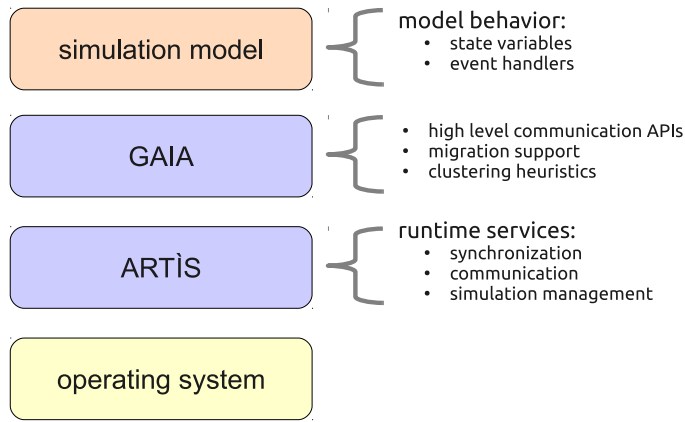}
\caption{High level structure of the simulator and its main components.}
\label{fig_logicscheme}
\end{figure}

The Generic Adaptive Interaction Architecture (GAIA) is a software layer built on top of the Advanced RTI System (ART\`IS) middleware~\cite{pads}. The high level structure of the simulation runtime is shown in Figure~\ref{fig_logicscheme}. ART\`IS is a PADS middleware that provides some services such as the simulation management (i.e.~LPs coordination, simulation bootstrap and shutdown, runtime statistics), the support for the main synchronization mechanisms (e.g.~conservative and optimistic) and the communication primitives for the interaction among LPs.

On top of the middleware, the GAIA framework exposes some higher level services to the simulation model. More in detail, GAIA provides the communication APIs (e.g.~interactions among SEs) and some basic utilities (e.g.~proximity detection) to speedup the model design and implementation. The main goal of GAIA is to implement the self-clustering mechanism (i.e.~model migrations and self-clustering heuristics) in a way that is transparent to the simulation model developer and correct under the simulation causality viewpoint. This is done implementing the migration protocol that is described in the following of this section and providing an easy way to migrate the state of SEs. In practice, GAIA implements the simulated model as a set of interacting SEs, where each SE is contained in a LP that is run on a PEU (as described in Section~\ref{sec:background}). As a consequence, the PADS is composed of a set of LPs (and its underlying PEUs) and the model evolution is obtained through the interactions among SEs. In practice, each SE is made of some state variables and a set of handlers triggered by the arrival of events (i.e.~interactions). These handlers are in charge of implementing the SE behavior.

It would have been possible to design GAIA in many different ways. In our view, the best approach is to see GAIA as a Multi-Agent Systems (MAS) in which each SE is an agent and the interactions among SEs are implemented as messages between agents. In the years, MAS have proven with good expressive power and easy to use.

In this paper, we are mainly interested in the GAIA adaptive model partitioning. Its main role is to analyze the interactions among SEs and to evaluate if and where some SEs should be clustered together. In other terms, it has to decide if the partitioning configuration has to be changed and provides the necessary support services.

Given the distributed nature of PADS, we think that it is not realistic to consider a centralized approach in which all the information is transfered to a hub where all the decisions about reallocations are taken. Therefore, in the following we will focus on self-clustering, that means that we are pursuing a decentralized scheme in which each LP is responsible for all SEs that it accommodates. In other words, the LP will analyze the interactions of its SEs and it will take decisions based on its findings and some inputs that it receives from the other LPs. Many different approaches can be followed to implement the SEs clustering described above. Given the partial information that is available to each LP,  the unpredictable nature of many parts of the system to simulate and of the simulator, we think that a good way is to rely on heuristics.

The LP could analyze the interaction patterns and measure the runtime conditions with sophisticated methods but this is often not profitable. It is necessary to remember that the reallocations goal is to reduce the Model Interaction Cost ($MIC$, Eq.~\ref{TotalExecutionCost-pads-exploded}), increasing the amount of local communications ($LCC$) while reducing the costly remote communications ($RCC$) (Eq.~\ref{ModelInteractionCost}). The other side of the dynamic reallocation is the Migration Cost ($MigC$, Eq.~\ref{MigrationCost}). As said before, the reallocation mechanism is profitable if and only if the cost paid for the new partitioning is lower than the saving given by the new configuration. Given that we are interested in systems composed of a very large number of parts, the $Heu$ (Eq.~\ref{MigrationCost}) could become a critical point. In other words, complex heuristics would introduce a high overhead when evaluating the interaction pattern of a large number of SEs. For this reason, in the following of this work we will concentrate on simple heuristics with a low computational cost and totally unaware of the simulation model semantics. 

A clustering heuristic designed for a specific simulation model would likely obtain better results than a general approach. The downside is that each type of simulated system would require a specific heuristic. We propose heuristics that are as general as possible, with the aim to study the outcomes of the proposed approach.

\subsection{Migration implementation}
\label{Migration}

\begin{figure}[ht]
\centering
\includegraphics[width=9.0cm]{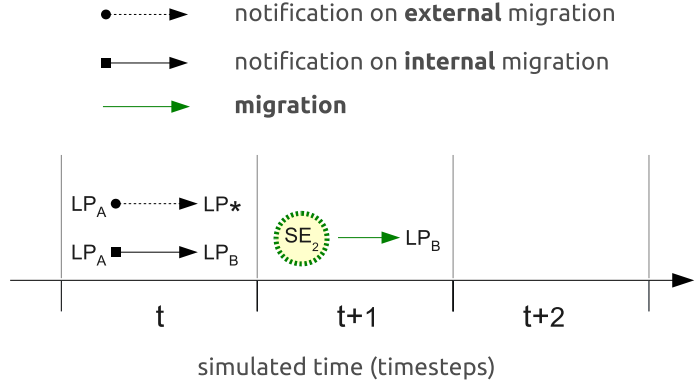}
\caption{Timing of the migration execution.}
\label{fig_migration_execution}
\end{figure}

Under a SE viewpoint, the LP is the provider of communication services, such as the interaction with other SEs. In practice, the real communication service is provided by the simulation middleware (in our case ART\`IS and GAIA). As seen above, in our implementation the LP is the container of a set of SEs that can flow along the execution architecture. In other words, the mapping between SEs and their containers (the LPs) can change during the simulation run. The consequence is that the middleware always needs an updated map of the SEs position in the execution architecture. In other words, every LP in the simulation has to be updated on the allocation of each SE. Talking about correctness, it is obvious that the dynamic reallocation of SEs can not alter, in any way, the semantic of the simulation. That is, the simulation based on adaptive partitioning must obtain the very same results than the one with static partitioning. Enabling a SE to freely move from a LP to another generates some problems, for example the synchronization between the LPs could be violated and some interaction could be lost. To avoid these issues, the migration has to be implemented following the procedure shown in Figure~\ref{fig_migration_execution} and described in the following.

As described above, the synchronization management is a key factor in PADS. Without synchronization, each PEU could run at its own speed and this would lead to the violation of the causal constraint between events. It is obvious that this would lead to meaningless simulation results. For the sake of simplicity, in this work we consider a time-stepped synchronization scheme in which the simulated time is divided in fixed-size timesteps and each LP can proceed to the next timestep only when all other LPs have completed the elaboration of the current timestep~\cite{1261535}. Anyway, the proposed adaptive partitioning can be adapted to work with other synchronization methods. For example, adding some synchronization barriers make it work with the Chandy-Misra-Briant synchronization scheme. The extension to an optimistic mechanisms (such as Time Warp) would be a bit more challenging but still possible. 

The decision of migrating some SEs is made by a self-clustering heuristic function that is implemented in each LP. The function is evaluated only on the local data of each SE and so, by design, can be seen as part of the SE itself. Let's suppose that at timestep $t$ the SE$_2$ (that is allocated in LP$_A$) finds that it should be migrated to LP$_B$. Before executing the migration, some extra actions have to be accomplished. First of all, LP$_A$ informs LP$_B$ that SE$_2$ will be migrated and that LP$_B$ is the destination (that is called ``notification of internal migration''). Due to the timestepped synchronization scheme, this message reaches LP$_B$ only at timestep $t+1$. For a correct simulation execution, also the other LPs (LP$_*$ in Figure~\ref{fig_migration_execution}) have to be informed of the migration. This is done using a so called ``notification of external migration'' message. More precisely, the first message is used by LP$_B$ for preparing the data structures. Instead, the second message is needed to inform the whole system that SE$_2$ will move and therefore, starting from timestep $t+1$, all messages with destination SE$_2$ have to be sent to LP$_B$. 

A constraint of the timestepped synchronization is that no message sent in timestep $t$ can be received in the same timestep, hence the minimum delivery time is timestep $t+1$. This constraint complicates the delivery of messages when migration in enabled. For example, multi-hop (or forwarding) schemes for the message delivery can not be used. In fact, a message delivered to a LP could find that the destination SE has just migrated away. In some circumstances, it would happen that there is no more time for delivering the message to the correct destination. In other words, when the event that is contained in the message should be executed, the message would be still in the wrong LP. Clearly this would lead to a causality violation (a synchronization error) that must be prevented. 

To avoid such problems, the migrations need to be considered in the messages delivery mechanism. Coming back to our example, at first in timestep $t+1$ the SE$_2$ executes all the events with timestamp equal to the current timestep and it produces some new events that have to be sent to other SEs. After that, the LP$_A$ serializes all the data structures of SE$_2$ and builds a ``migration message'' to move the SE from LP$_A$ to LP$_B$. Such a message, sent at timestep $t+1$, arrives in LP$_B$ at timestep $t+2$. The receiving LP, at first of timestep $t+2$, collects all the received ``migration messages'' and creates the newly arrived SEs. In the following of the timestep, the migrated SEs are able to process their events for the current timestep as usual. In fact, the messages containing such events were sent by the other LPs directly to LP$_B$, since that was set as the migration destination by the previous ``notification of external migration'' message.

Another interesting aspect is about the delivery, at timestep $t$, of events with timestamps that are larger than $t+1$ (that is, $t+\delta$). The delivery of such kind of messages can be implemented in different ways. A first choice is to deliver the events to the destination LP as soon as they are sent. This means that the destination LP (that is the LP that contains the destination SE) could have to store such messages for a given amount of timesteps. The problem is that, in our design the SEs can be migrated, and therefore it is not possible to know ``a priori'' what LP will contain the destination SE at the event timestamp. In other words, in the timesteps that are from the sending of the message ($t$) and its timestamp ($t+\delta$), the SE could be migrated many times and this approach would require also to relocate all the messages having such SE as destination. Another simpler approach is to store the sent messages in the originating LP up to their delivery.
This means that an event with timestamp $t+\delta$ is stored in the originating LP up to the timestep $t+\delta-1$. Only at such timestep, the message is sent to the LP that will contain the destination SE in the next timestep. All this can be easily implemented using the previously described mechanism for managing migrations. In this way, even if the destination SE has changed many LPs (due to multiple migrations), a single message delivery is sufficient. Furthermore, following this approach, the events sent by a SE are not part of its local state and therefore, in case of migration, have not to be serialized and transfered on the network. The advantages of this approach are quite clear: it is simple, minimizes the SEs state size and avoids multiple retransmissions of messages. On the other hand, the main drawback is the need to store events in the originating LP for a (potentially) long time. In our view, this is acceptable given that, in any case, such messages have to be stored in some part of the execution architecture. Another minor issue is that all LPs have to stay in the simulation for its whole length. Otherwise, the events that are stored in the LP that is leaving the simulation would be lost. In practice, this constraint is usual in many conservative synchronization approaches. Moreover, the implementation of simulations with a dynamic number of LPs can be done adding a simple procedure: before exiting the simulation the LPs have to deliver all the stored messages to a randomly chosen LP (that is not leaving). Such destination LP will be responsible for the final delivery of messages when appropriate.

Now that the design of the migration mechanism has been described, we can focus on the self-clustering heuristics.

\subsection{Self-clustering heuristics}
\label{Heuristics}

The goal is to define a self-clustering heuristic that is able to partition the SEs in a given set of groups while minimizing the amount of inter-group (i.e.~inter-LP ) communication and the number of migrations. The heuristic functions are going to be evaluated in each LP and must rely only on local information and avoid every kind of centralization. All of this has to be done using as little computation as possible, for better performance and scalability. This last point is very crucial given that many simulated systems are often modeled by a huge number of SEs and each of them has to be analyzed by the clustering heuristic. Due to the constraint of using only local information and the partial view of the system of each LP, all the self-clustering heuristics considered in this paper are based on the analysis of the communication pattern of SEs (e.g.~amount of intra-LP vs. inter-LP communication). That is an information locally available to each LP and easy to process.

As said before, we are not interested in specific solutions for some kind of simulation models, our aim is for a general approach. On the other hand, in practice, there are many different kinds of simulated systems, each one with characteristics and issues that need to be addressed. For this reason, GAIA implements a small set of heuristics and each of them implements a slightly different evaluation mechanism. This allows to match each simulated system with the more appropriate clustering heuristic. Furthermore, each of the heuristics can be fine-tuned using some runtime parameters that are exposed to the simulation model. With some extra effort, it would be possible to analyze the characteristics of the simulated model at runtime and to trigger the more appropriate clustering heuristic (see Section~\ref{SelfTuningAdaptPartitioning}). Anyway this aspect is out of the scope of this paper.

\subsubsection{Heuristic \#1}
\label{Heuristic_1}

This heuristic is evaluated at each timestep and considers the last $\kappa$ simulation timesteps. For each SE in the LP, a sliding window mechanism is used for accounting its interactions. The core function of the mechanism is to find which LP (excluding the local one) has been the destination of the most part of the interactions sent by the SE. The total amount of such ``external interactions'' (during the last $\kappa$ timesteps, to this specific LP) is called $\epsilon$. This value is compared with the amount of ``internal interactions'' ($\iota$) sent during the same time frame to SEs that are allocated on the same LP.

\begin{equation}
\label{MigrationFactor}
	\alpha = \frac{\epsilon}{\iota}
\end{equation}

At this point, the SE is tagged as a ``candidate for migration'' only if : i) $\alpha > MF$ where MF (Migration Factor) is a parameter that is used to control the number of migrations in the system, and ii) at least MT (Migration Threshold) timesteps have passed since the last migration of this specific SE.

The rationale behind this heuristic is to use a time window to evaluate the communication pattern of each SE. A SE does not not need to be migrated if the most part of its interactions are delivered to its current LP. In case there is another LP that is the destination of most part of the SE interactions then a migration may be necessary. That migration will happen only if the unbalance in the amount of communication is large enough to justify the cost of a SE migration (i.e.~the MF parameter). It is well known that this kind of systems could exhibit an oscillatory behavior. For this reason, the MT parameter has been added. This means that, a SE is not allowed to migrate to a different LP in each timestep. This reduces the ability of the self-clustering mechanism to adapt to very dynamic environments but also prevents some extreme behaviors that would be hard to control.

This clustering heuristic is directed to systems with a lot of communication among SEs and where empty timesteps are quite rare. The $\kappa$ parameter, used to define the sliding window size, controls the importance of new events with respect to old ones. In other words, a sliding window that is very small will likely overreact to new behaviors, increasing the number of requested migrations. Conversely, a too large one will be unable to react to new trends. Given that this heuristic has to be evaluated for each SE at every timestep, if the timestep length is quite small (e.g. for the modeling of medium access control protocols) the resulting overhead could be quite large.

\subsubsection{Heuristic \#2}
\label{Heuristic_2}

This heuristic, such as the previous one, is evaluated at each timestep. Each SE has an associated sliding window in which the last $\omega$ interactions sent by the SE are stored. In this way, each time a new event is added to the window then the oldest one is discarded. The effect of this mechanism is that if a SE has a low interaction rate, then the heuristic is still able to consider the old events. This would not happen with Heuristic \#1 since it only considers the last $\kappa$ simulation timesteps. The rest of the mechanism is the same as in Heuristic \#1: the $\alpha$ ratio is calculated and compared to MF. Again, the MT parameter can be used to pace the migration rate.

The rationale behind this heuristic is the same of Heuristic \#1 but with a different management of the data used for evaluating the clustering heuristic. Instead of assuming a fixed-size time window associated to each SE, in this case, a fixed number of interactions for each SE are evaluated. The difference is subtle, but it can lead to significant differences in some systems. In fact, in the systems in which the generation of new events is rare then the Heuristic \#1 would not find SEs to migrate (i.e.~the time window would be almost empty). Conversely, the analysis of the last interactions generated by each SE could lead to some clustering.

\subsubsection{Heuristic \#3}
\label{Heuristic_3}

In models with a large number of SEs, the cost of evaluating the heuristic, at every timestep for every SE, can be prohibitive. This heuristic implements the same function of Heuristic \#2 but in which the evaluation is triggered only if the analyzed SE has sent at least $\zeta$ interactions since the last evaluation. In many systems, this approach permits to greatly reduce the number of evaluations at each timestep and therefore improve the scalability of the mechanism. The rest of the heuristic is exactly the same as in \#2 and therefore it can be tuned using the MF and MT parameters.

The clustering obtained by heuristics \#2 and \#3 is very close but the difference in terms of evaluation cost can be large. In fact, in some systems, the ability to skip all the SEs that are silent can lead to a significant reduction in the computational cost.

\subsubsection{Comparison of Heuristics}
\label{Heuristic_comparison}

The performance comparison of the previously described self-clustering heuristic is not among the goals of this paper since we are interested in showing that even a generic (and simple) heuristic can obtain good results. Under the design viewpoint, it is worth noting that all the heuristic described above follow the same self-clustering approach (that is based on the ratio between the amount of remote and local communications). Despite of this, there are some small differences in the design of the heuristics that can led to big differences both in the overhead of computing the heuristics and on the clustering performances.

In the scientific literature on MAS and data mining many different clustering approaches have been proposed~\cite{Ogston:2003:MDC:860575.860702}. It worth noting that a clustering heuristic to be used in GAIA must have some specific characteristics. For example, for cost reasons, the evaluation must be based only on a local view of the system (i.e.~all the data used by the heuristic must be already available in the LP in which the SE is allocated). Moreover, in our case, the main goal of clustering is the minimization of the inter-LP communication. This means that, many interesting approaches to clustering in MAS can not be easily implemented in GAIA or that would require large changes. More in detail, many variants of the k-means clustering~\cite{Jain:1988:ACD:42779} have been proposed but the implementation of most of them requires a global knowledge of the MAS~\cite{Ogston:2003:MDC:860575.860702}. Hierarchical approaches~\cite{Guha:1998:CEC:276305.276312,5299196}, k-medoid algorithms~\cite{Ng:1994:EEC:645920.672827} and the different form of swapping described in~\cite{6483388} have the same issue~\cite{Ogston:2003:MDC:860575.860702}. All the clustering algorithms based on a sample set~\cite{KaufmanR90}, an approach often used to enhance scalability, can not be used in GAIA since all the SE must be evaluated and allocated in a specific LP. An interesting approach is described in~\cite{Judd:1998:LPD:284980.284991} in which there is a local pre-clustering and in the following the combination of the local results. In~\cite{Chaimontree2012}, an initial cluster configuration is generated, in the following firstly there is 
a bidding phase among the agents and secondly a final phase in which the clustering agents negotiate with one another so as to improve the initial cluster configuration. The main drawback of this kind of two-phases procedures is that, the presence of multiple phases adds some delay in the clustering execution.

\subsection{Load Balancing}
\label{subsec:loadbalancing}

As described in Section~\ref{sec:introduction}, the partitioning of the simulated model involves both communication and computation aspects. In fact, if only communication is considered then the best strategy would be to cluster all the SEs in the same LP. In this way, the Model Interaction Cost ($MIC$, see Eq.~\ref{ModelInteractionCost}) is equal to the $LCC$ and the $RCC$ is zero. In other words, it would be a monolithic simulation in which a single PEU executes all the model computation (i.e.~$f(N)=1$ in Equation~\ref{TotalExecutionCost-pads}). A more realistic approach is to adapt the partitioning with the aim to reduce the $MIC$ while preserving the parallelization of computation. In practice, this means that the outcomes of the clustering heuristic have to be constrained by a symmetric or asymmetric load balancing scheme.

In symmetric load balancing, the underlying assumption is that each PEU has the proper computation and communication load for not being the bottleneck of the execution architecture. In other words, the execution architecture is well balanced and the load balancing mechanism must not introduce imbalances caused by SEs migrations. Considering a single LP, and assuming that SEs are quite homogeneous, the load balancing scheme is symmetric if the number of outbound migrations equals the inbound ones. As usual, this approach can be implemented in many different ways. In GAIA, the outcomes of the clustering heuristic are scrutinized by a load balancing mechanism that forbids the migrations that would cause imbalances and allows all the others. More in detail, if we assume that the clustering heuristic is evaluated at timestep $t$, then at the end of this timestep each LP broadcasts to all other LPs the total number of SEs that should be migrated to each specific destination (i.e.~candidates for migration). In the next timestep, $t+1$, each LP has collected all the data for balancing the inbound and outbound migrations. Now, each destination LP can communicate to each source LP the exact number of migrations that it can accept from that source. Finally, at timestep $t+2$ the real migration procedure can start (as illustrated in Section~\ref{Migration}). The main drawback of this implementation is the further delay that is added to the migration mechanism. In fact, the amount of time from the triggering of a migration and its completion is of a few timesteps (2 for the load balancing scheme and 3 for the migration procedure). In practice, this delay can affect the performance of the adaptive self-clustering mechanism (that is a reactive system).

In asymmetric load balancing, the inbound and outbound migrations, in a specific LP, may be imbalanced. This means that this scheme would be able to deal with PEUs with different execution speed, to react to the presence of background communication/computational load and to adapt to the runtime characteristics of the simulation model. In this case, each LP can accommodate a specific number of SEs, that is not fixed, and that can be influenced by many factors. For example, a PEU equipped with a very fast CPU would run a larger number of SEs than slower PEUs, up to when it starts some background load (e.g. a process unrelated with the simulation is executed on the same CPU). In asymmetric load balancing the imbalances are not only permitted but caused by the load balancing mechanism to adapt the model partitioning to the runtime characteristics of the execution architecture. More in detail, instrumenting the synchronization mechanism that is implemented in ART\`IS, it is possible to collect live data on the simulation execution. This data is then used to allow imbalanced migrations from the slow LPs, that can be forced to reduce the number of allocated SEs (in fact reducing their load), to faster LPs (hence slowed down). In practice, this approach permits a smooth execution in which bottlenecks are mitigated.

Our simulation middleware supports both symmetric and asymmetric load balancing but in this paper we will consider only the symmetric case. This because, in this case, our main goal is to test the proposed adaptive clustering approach and to assess the performance of the communication-based clustering heuristics. Hence, the asymmetric load balancing would have led to a more complex analysis without any advantage. The testbed that results from our choice, and that will be used in the following performance evaluation, is still realistic and relevant since it is a typical High Performance Computing (HPC) execution environment. In most HPC systems, all the computation nodes are homogeneous in terms of performance and there is no background load that produces communication and computation imbalances.

\section{Performance Evaluation}
\label{Performance}

Due to the characteristic of the partitioning problem (and the systems to simulate) it is not possible to find analytically what is the best allocation of the SEs (on the LPs) for each timestep. As a consequence, it is not possible to demonstrate the general validity of the self-clustering approach: in some cases this approach leads to a performance gain, in others to a loss. 

For these reasons, a simulation model has been defined (and implemented) with the aim to evaluate the performance of the adaptive partitioning using a use case. The choice of the model to use is not easy. It must be general enough to be representative of a wide class of simulation models but sufficiently specific to provide realistic results. Furthermore, it must not represent a best/worst case for the self-clustering mechanism or the performance evaluation would produce misleading results.

\subsection{Simulated model}
\label{SimModel}

The model of choice is an Agent-Based Model~\cite{Macal:2005:TAM:1162708.1162712} in which a bidimensional space (with no obstacles) is populated by a given number of agents. The simulated area is toroidal and every agent moves following a defined mobility model. The interaction among agents is based on proximity. In other words, when an agent produces a new interaction, this will be delivered to all agents that are within a threshold distance. Varying the mobility model and the interaction characteristics, it is possible to evaluate a wide range of conditions. In its main characteristics, the model is an abstraction of a wireless ad hoc network.  

Applying the adaptive self-clustering to this model is straightforward. In fact, each agent can be represented by a SE and the interactions among agents can be mapped as communications among SEs. As said before, the goal of the adaptive self-clustering is to migrate the SEs with the aim to reduce the communication cost in PADS. Clearly, the benefit of clustering is counteracted by the agents movement, that causes an ever changing interaction pattern among the SEs. For this reason, the choice of the mobility model is crucial for evaluating the performance of the mechanism. For example, a mobility model that preserves the locality of agents would give a boost to the clustering. In other terms, if the agents follow a group mobility model in which they change their absolute position in the simulated area but not their relative position, then it would be easy to obtain a stable clustering of the interacting SEs.

To comply with that requirements, the mobility model of choice is the Random Waypoint (RWP)~\cite{rwp}. It is one of the most popular mobility models, in which every agent is free to move in the whole simulated area and there is no correlation among the agents movement. For low movement speeds, it allows a certain level of locality in the communication among agents, that is disregarded with higher speeds. Furthermore, it is not a mobility model in which the nodes do short walks around their initial position. In our opinion, the RWP is a balanced choice. In fact, in the real world, roads and obstacles have the effect to bound the agents movement and therefore to force a higher amount of locality in the interaction pattern. On the other hand, in most cases, even a simulation model in which the agents are not in a physical space would show some degree of time locality in the interaction among agents.

\subsection{Self-clustering}
\label{SelfClustering}

The first step in the performance evaluation is to verify whether the adaptive self-clustering mechanism, in the prosed simulation model, is able or not to cluster the interacting SEs and to maintain a good level of clustering for the whole simulation. In this case, the clustering efficiency is evaluated through an indirect measure, the Local Communication Ratio (LCR). The LCR is the ratio of the number of ``local'' (intra-LP) communications by all SEs in a given LP, with respect to the total amount of interactions originating from this LP. In other words, it represents the partitioning of MIC (as seen in Eq.~\ref{ModelInteractionCost}). It is a coarse grained measure that does not take in account many details, that will be needed in the following, but that is useful to measure the amount of clustering that can be obtained in different models or conditions. Clearly, the clustering of SEs is obtained at the cost of a certain number of migrations. We can expect that the more dynamic the simulated scenario (e.g.~higher movement speed), the more migrations will be needed to obtain the same LCR level.

\subsubsection*{Experiment 1} In this first experiment, the number of SEs (\#SE) is set to $10000$ and the number of LPs (\#LP) is $4$. Each SE is randomly placed in the simulated area and assigned to a LP. The assignment of SEs to the LPs is random but each LP gets the same number of SEs. The mobility model is RWP with minimum speed equal to maximum speed and in the range $[1,29]$ (spaceunits/timestep), the sleep time of RWP is set to 0 timesteps. At every timestep the $100\%$ of nodes is in motion. The simulated area is a toroidal square of $10000x10000$ spaceunits and the simulation length is set to $3600$ timesteps. The threshold interaction range is $250$ spaceunits and the probability, for a SE to send an interaction in a given timestep, is set to $0.2$. This means that at each timestep the $20\%$ of SEs sends interactions. Finally, the adaptive self-clustering heuristic \#1 is used with tuning parameters: MF in the range [$1$, $20$] and MT=$10$. The performance evaluation shown in this paper is based on heuristic \#1 while a more detailed comparison of the heuristics is left as future work.

\begin{figure}[h]
\centering
\includegraphics[width=12.0cm]{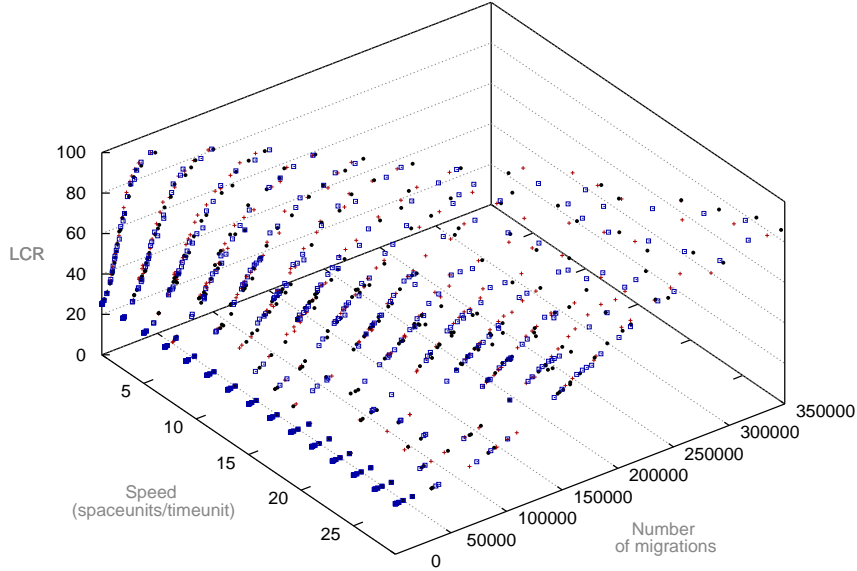}
\caption{Effect of mobility on the self-clustering performance. Local Communication Ratio (LCR) that can be obtained with a given speed and with an increasing number of migrations. Multiple independent runs (points in different colors and shapes).}
\label{fig_perf_1}
\end{figure}

For every speed value and for every MF in the studied ranges, a new simulation run has been executed. The resulting average LCR and the related number of migrations are shown in Figure~\ref{fig_perf_1}. Even if, in this case, our only goal is to see if there are visible trends (and not to perform a quantitative evaluation), multiple independent runs have been executed (different point styles and colors in the figure). The main trend is quite clear: for low speed values a very limited number of migrations is enough for obtaining very high LCR values. It worth noting that the LCR value that can be obtained in a PADS with $4$ LPs and a static allocation of SEs in the LPs is $25\%$. In a moderately dynamic system, very few migrations are needed to rearrange the whole PADS allocation and to bump the LCR up to $90\%$. With higher speed values, it is still possible to obtain a good level of clustering, but the number of required migrations is higher and higher. In other words, the decrease in time/space locality forces the self-clustering heuristic to more frequent migrations.

\subsubsection*{Experiment 2} 
\begin{figure}[h]
\centering
\includegraphics[width=8.5cm]{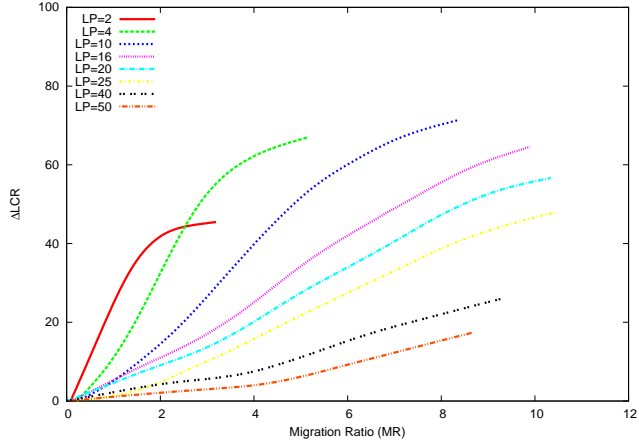}
\caption{Effect of the number of LPs on the self-clustering performance. Multiple independent runs, averaged results.}
\label{fig_perf_2}
\end{figure}

The number of LPs in which a model is partitioned has an impact on the simulator performance (see Eq.~\ref{TotalExecutionCost-pads-exploded}) and on the efficiency of the self-clustering mechanism. With respect to the LCR, partitioning the simulation in more and more LPs reduces the probability that the interaction between a pair of SEs is local to a single LP. In fact, each LP allocates a smaller percentage of the SEs in the simulation. In this experiment, we investigate the $\Delta$LCR with respect to the Migration Ratio (MR) when the number of LPs is in the range [$2$, $50$]. The $\Delta$LCR is defined as the difference between the average LCR with and without the self-clustering mechanism. A positive value means that the mechanism is able to cluster the interacting SEs (higher is better). Zero or negative means that the self-clustering mechanism is adding overhead without being able to cluster the interacting SEs (see Eq.~\ref{MigrationCost}). As seen in the previous experiment, the higher the number of migrations,  the higher the LCR (and $\Delta$LCR) that can be obtained. Unfortunately, the number of migrations is not a good metric because it depends on factors such as the number of SEs and the simulation length. In other words, we need a metric that is comparable between different simulation configurations. The MR is defined in Equation~\ref{MigrationRatio}, where \#SEs is the total number of SEs in the simulation and the simulation length is expressed in timesteps. All the other parameters of this experiment are the same of Experiment 1, except the speed that, in this case, is set to $11$ spaceunits/timestep. The results are show in Figure~\ref{fig_perf_2} and demonstrate that if the number of LPs in the simulation is quite moderate then large LCR gains can be obtained. When the simulation is divided in larger number of parts, it is more difficult to cluster the interacting SEs but there is still some gain. It worth noting that, in this experiment, the problem size is not changed (e.g.~the number of SEs) while it is partitioned in more and more parts.

\begin{equation}
\label{MigrationRatio}
	MR = \frac{TotalNumberOfMigrations}{\#SEs(\frac{SimulationLength}{1000})}
\end{equation}

\subsubsection*{Experiment 3} The threshold interaction range influences the communication behavior of SEs. The larger the range, the more the SEs that receive each interaction. In this experiment, all the parameters are set as before (e.g.~speed=$11$ spaceunits/timestep) but the number of LPs is 4 and the interaction range is varied in the set \{50, 100, 200, 400, 800, 1600\} spaceunits.

\begin{figure}[h]
\centering
\includegraphics[width=8.5cm]{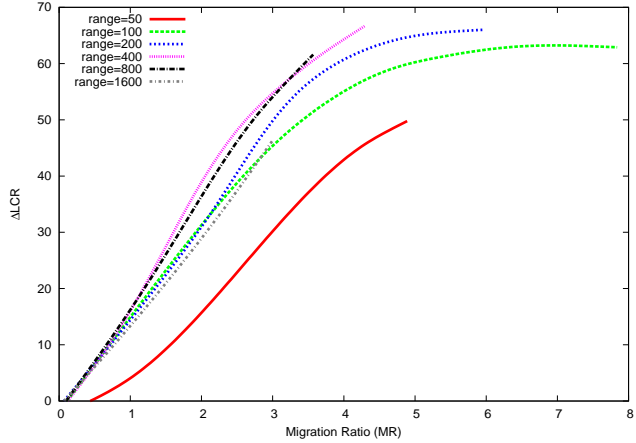}
\caption{Effect of the threshold interaction range on the self-clustering performance. Multiple independent runs, averaged results.}
\label{fig_perf_4}
\end{figure}

The effect of the interaction range on the $\Delta$LCR is shown in Figure~\ref{fig_perf_4}. When the range is small (e.g.~$50$ spaceunits) the self-clustering generates a high number of small clusters that are prone to disaggregation. In fact, the range is so small that a few timesteps are enough to move the SEs from a given set of interacting neighbors to another. The effect of this behavior on the self-clustering performance is a large number of useless migrations (i.e.~a high MR and medium $\Delta$LCR). Increasing the interaction range (e.g.~$100$, $200$) has the effect to stabilize the clusters and to improve the clustering performance. This behavior continues up to a tipping point ($400$ in this specific simulation model). For larger interaction ranges there is a degradation in the performance (e.g.~$800$ and $1600$). The effect of a large interaction range is that each interaction is received by many SEs and therefore the interaction sets of SEs are overlapped. In other terms, each SE has too many neighbors. In terms of performance, this makes difficult to maintain a good clustering.

\subsection{Migration cost}
\label{SubSec:MigrationCost}

The adaptive partitioning has a cost structure that has been described in Equation~\ref{MigrationCost} ($MigC$). The cost of communication in parallel, LAN-based and Internet-based execution grows according to the distance between the computational units in the execution architecture. For example, the network latency in parallel execution environments is orders of magnitude lower than on Internet. Given that, in most cases, the communication is the most relevant factor in $MigC$ then this cost will be strongly affected by the latency and other parameters such as the state size (e.g. the amount of state variables) in the SEs to be migrated. It is straightforward to demonstrate that $MigC$ increases when the network latency gets higher and that the state size has a big impact when the network bandwidth is limited (e.g.~in Internet-based execution environments).

To enable the assessment of $MigC$, GAIA/ART\`IS has been slightly modified to work as follows. The SEs interactions are handled as usual from the model viewpoint (e.g.~for the self-clustering heuristic) but in practice, all the interactions never produce any network load (e.g.~communication among the LPs). This means that, in this configuration of the runtime, the only communication in the simulation is for the LPs synchronization ($SC$ in Eq.~\ref{TotalExecutionCost-pads-exploded}) and the migration of SEs. This means that, in this case, to assess $MigC$ is sufficient to compare the simulator execution time (that is Wall Clock Time, WCT~\cite{fujimoto99}) with and without the adaptive self-clustering mechanism. Hence, the difference between the two execution times is the measure of the overhead that is needed for computing the self-clustering heuristic, serializing the SEs data structures and transferring their state. For conciseness and clarity, we have decided not to report a complete evaluation of $MigC$ and to give more space to the adaptive partitioning evaluation (that is reported in the following of this section).

\subsection{Adaptive partitioning}
\label{AdaptPartitioning}

In the last part of this section, we investigate if the adaptive partitioning mechanism is able to offer a speedup in the execution of parallel/distributed simulations. In other words, two different execution environments (i.e. parallel and distributed) will be used and, for each of them, the simulator execution time (i.e.~WCT) with and without the self-clustering will be compared (in the following of this section this measure will be called $\Delta$WCT).

The simulation model used in this evaluation is defined in Section~\ref{SelfClustering} (Experiment 1). More specifically, the number of LPs is set to $4$ and the total number of SEs is $10000$. At every timestep, all the nodes are in motion, the minimum speed is equal to maximum speed and is set to $11$ (spaceunits/timestep). Moreover, the sleep time is disabled and the interaction range is set to $250$ spaceunits. In this case, the simulation length has been set to $1200$ timesteps in order to limit the amount of computing time required for running all the different setups. In the past, we have already demonstrated that, when in appropriate conditions, the self-clustering mechanism implemented can speedup simulations~\cite{gda-ijspm-2009}. In this paper, we are interested in the application of the cost analysis that we have introduced to a simulation model that is not favorable (i.e.~all SEs move at each timestep, the speed is not moderate and with very little space locality). The three main parameters of this performance evaluation are: i) the characteristics of the simulation execution architectures (i.e.~parallel vs. distributed), ii) the state size of SEs (to be migrated by the self-clustering mechanism) and iii) the amount and size of the interactions that are exchanged during the simulation by the SEs. The execution architectures considered are: a parallel (multi-core) computer and a distributed LAN-based cluster. In future work, we will investigate the performance in distributed Internet-based clusters (e.g.~multicloud environments). The state size of SEs has been varied in the set \{$32$, $20480$, $81920$\} bytes: $32$ is the state size of SEs in the simulation model that has been used; $20480$ and $81920$ are obtained adding some padding. The size of each interaction has been studied in the set \{$1$, $100$, $1024$\} bytes: $1$ is the minimum size allowed by this specific simulation model; the other configurations are obtained adding some padding to the interaction messages. The amount of interactions delivered in each timestep has been varied setting to $0.2$ and $0.5$ the probability ($\pi$) that, in a given timestep a given SE sends an interaction (that is $20\%$ and $50\%$).

The parallel setup is a DELL R620 with 2 CPUs and 128 GB of RAM. Each CPU is a Xeon E-2640v2, 2 GHz, 8 physical cores. Each CPU core is with hyper-threading enabled and therefore the total number of logical cores is 32. The computer is equipped with Ubuntu 14.04.3 LTS and GAIA/ART\`IS version 2.1.0. The distributed setup is a cluster of servers interconnected by a Gigabit Ethernet LAN. The specifications of each node in the cluster are reported in Table~\ref{table_distributed_specs}.

\begin{table}[h]
\begin{center}
	\begin{tabular}{ | c | c | c | c | c | }
	\hline
	\textbf{CPU} & \textbf{\shortstack{Physical \\ Cores}} & \textbf{\shortstack{Hyper \\ Threading}} & \textbf{RAM} & \textbf{\shortstack{Operating \\ System}} \\ \hline
	Intel Xeon X3220 2.40 GHz & 4 & No & 4 GB & \multirow{4}*{Ubuntu 12.04.5 LTS} \\
  Dual-Intel Xeon 2.80 GHz & 2x1 & Yes & 3 GB & \\
  Dual-Intel Xeon 2.80 GHz & 2x1 & Yes & 4 GB & \\
  Dual-Intel Xeon 2.80 GHz & 2x1 & Yes & 2 GB & \\ \hline 
	\end{tabular}
\end{center}
\caption{Specifications of the distributed setup.}
\label{table_distributed_specs}
\end{table}

All the results reported in this section are averages of multiple independent runs. In all cases, the confidence intervals ($90\%$) have been calculated but not reported for readability. In both parallel and distributed setups, the widths of the confidence intervals are less than $1.71\%$ of the mean values.

In Table~\ref{table:parallel}, the results for the parallel setup are reported. Every configuration, in terms of Migration Size and Interaction Size, has been tested with both moderate and high dissemination probability (i.e.~$\pi$). In all cases, firstly it has been measured the WCT required to complete a simulation run with static allocation of SEs (i.e.~GAIA OFF). Secondly, this result has been compared with the best WCT that can be obtained with GAIA ON. This difference is the gain/loss provided by the self-clustering scheme (i.e.~$\Delta$WCT). To find the best WCT with GAIA ON, the whole MF range [$1.1$, $19$] has been explored. The configurations in which there are the worst and the best results are reported in bold. To improve readability, the WCT reduction obtained by GAIA is reported as a positive number. Conversely, when GAIA slows down the execution this is reported as a negative number. In term of performance, the communications in the parallel setup have very high bandwidth and very low latency. This means that both $MIC$ (Eq.~\ref{TotalExecutionCost-pads-exploded}) and $MigComm$ (Eq.~\ref{MigrationCost}) have a lower impact on $TEC$ (see Eq.~\ref{TotalExecutionCost-pads}) with respect to setups in which the communications are more costly (e.g.~LAN or Internet based). If the communications, in the parallel setup, have a limited impact on $TEC$, then the main cost factor must be given by the computation. In this paper, only the communication load balancing features of GAIA are evaluated. This means that, in a scenario in which computation is the predominant cost, very large gains can not be expected from a mechanism that acts only on communication.

For all the reported configurations, GAIA is able to obtain a gain that is $1.67$\% in the worst case and $19.47$\% in the best one. The worst configuration is when Interaction Size is $1$ and Migration Size is $81920$. This means that, even if the self-clustering is able to save ``costly'' $RCC$ for ``cheaper'' $LCC$ (Eq.~\ref{TotalExecutionCostWithMigration}), the cost paid for migrations ($MigC$ in Eq.~\ref{MigrationCost}) is so high that the gain is narrow. On the opposite, the best configuration is with large Interaction Size (e.g.~$1024$) and small Migration Size (e.g.~$32$). It is worth noting that, in all the reported cases, the best results have been obtained tuning GAIA for an aggressive behavior (i.e.~a large number of migrations, that is a low MF in Heuristic \#1). Increasing the interaction probability ($\pi$) has the effect to increase the relative amount of $CC$ in $TEC$ (Eq.~\ref{TotalExecutionCost-pads}). The effect is that the enhancement provided by GAIA has a larger effect on $TEC$.

\begin{table}[]
\centering
\begin{tabular}{ccc|cc|cc|}
\cline{4-7}
                           &                                     &                  & \multicolumn{2}{c|}{$\pi=0.2$}         & \multicolumn{2}{c|}{$\pi=0.5$}         \\ \hline
\multicolumn{1}{|l|}{GAIA} & \multicolumn{1}{l|}{Migr.~Size} & Inter.~Size & \multicolumn{1}{l|}{WCT (MF)} & $\Delta$WCT & \multicolumn{1}{l|}{WCT (MF)} & $\Delta$WCT \\ \hline
\multicolumn{1}{|c}{OFF} & \multicolumn{1}{c}{-} & \multicolumn{1}{c}{1} & \multicolumn{1}{|c}{94.87} & \multicolumn{1}{c|}{-} &\multicolumn{1}{|c}{204.97} & \multicolumn{1}{c|}{-} \\
\multicolumn{1}{|c}{ON} & \multicolumn{1}{c}{32} & \multicolumn{1}{c}{1} & \multicolumn{1}{|c}{89.70 (1.2)} & \multicolumn{1}{c|}{5.45\%} & \multicolumn{1}{|c}{187.68 (1.2)} & \multicolumn{1}{c|}{8.44\%} \\
\multicolumn{1}{|c}{ON} & \multicolumn{1}{c}{20480} & \multicolumn{1}{c}{1} & \multicolumn{1}{|c}{90.80 (1.2)} & \multicolumn{1}{c|}{4.30\%} & \multicolumn{1}{|c}{191.24 (1.1)} & \multicolumn{1}{c|}{6.70\%} \\
\multicolumn{1}{|c}{ON} & \multicolumn{1}{c}{81920} & \multicolumn{1}{c}{1} & \multicolumn{1}{|c}{\bf 93.29 (1.2)} & \multicolumn{1}{c|}{\bf 1.67\%} & \multicolumn{1}{|c}{195.44 (1.5)} & \multicolumn{1}{c|}{4.65\%} \\ \hline
\multicolumn{1}{|c}{OFF} & \multicolumn{1}{c}{-} & \multicolumn{1}{c}{100} & \multicolumn{1}{|c}{98.48} & \multicolumn{1}{c|}{-} &\multicolumn{1}{|c}{208.69} & \multicolumn{1}{c|}{-} \\
\multicolumn{1}{|c}{ON} & \multicolumn{1}{c}{32} & \multicolumn{1}{c}{100} & \multicolumn{1}{|c}{91.14 (1.3)} & \multicolumn{1}{c|}{7.45\%} & \multicolumn{1}{|c}{194.81 (1.6)} & \multicolumn{1}{c|}{6.65\%} \\
\multicolumn{1}{|c}{ON} & \multicolumn{1}{c}{20480} & \multicolumn{1}{c}{100} & \multicolumn{1}{|c}{93.36 (1.1)} & \multicolumn{1}{c|}{5.20\%} & \multicolumn{1}{|c}{197.14 (1.5)} & \multicolumn{1}{c|}{5.54\%} \\
\multicolumn{1}{|c}{ON} & \multicolumn{1}{c}{81920} & \multicolumn{1}{c}{100} & \multicolumn{1}{|c}{95.46 (1.2)} & \multicolumn{1}{c|}{3.07\%} & \multicolumn{1}{|c}{198.83 (1.1)} & \multicolumn{1}{c|}{4.72\%} \\ \hline
\multicolumn{1}{|c}{OFF} & \multicolumn{1}{c}{-} & \multicolumn{1}{c}{1024} & \multicolumn{1}{|c}{130.11} & \multicolumn{1}{c|}{-} &\multicolumn{1}{|c}{288.92} & \multicolumn{1}{c|}{-} \\
\multicolumn{1}{|c}{ON} & \multicolumn{1}{c}{32} & \multicolumn{1}{c}{1024} & \multicolumn{1}{|c}{109.02 (1.3)} & \multicolumn{1}{c|}{16.22\%} & \multicolumn{1}{|c}{\bf 232.66 (1.1)} & \multicolumn{1}{c|}{\bf 19.47\%} \\
\multicolumn{1}{|c}{ON} & \multicolumn{1}{c}{20480} & \multicolumn{1}{c}{1024} & \multicolumn{1}{|c}{110.97 (1.1)} & \multicolumn{1}{c|}{14.71\%} & \multicolumn{1}{|c}{234.24 (1.1)} & \multicolumn{1}{c|}{18.92\%} \\
\multicolumn{1}{|c}{ON} & \multicolumn{1}{c}{81920} & \multicolumn{1}{c}{1024} & \multicolumn{1}{|c}{112.22 (1.2)} & \multicolumn{1}{c|}{13.75\%} & \multicolumn{1}{|c}{235.75 (1.3)} & \multicolumn{1}{c|}{18.40\%} \\ \hline
\end{tabular}
\caption{Parallel setup with self-clustering mechanism OFF/ON. Different migration and interaction sizes. Different probabilities ($\pi$) to generate new interactions. The results reported are: the mean WCT (in seconds), the Migration Factor (MF) used to tune GAIA and the gain/loss obtained by self-clustering (percentage $\Delta$WCT). In bold, the worst and best performance obtained in the parallel setup.}
\label{table:parallel}
\end{table}

\begin{figure}[h]
\centering
\includegraphics[width=8.5cm]{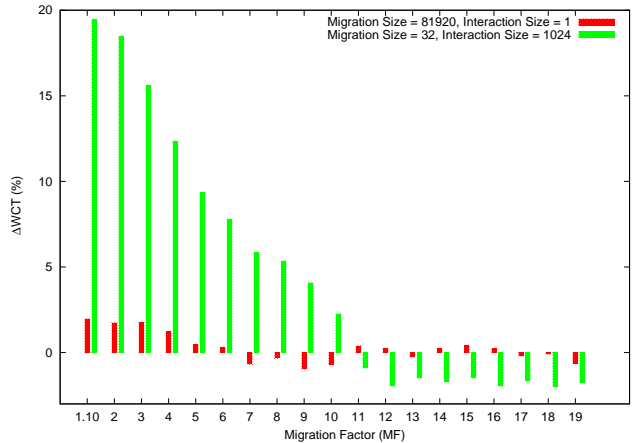}
\caption{Parallel setup: best (green) and worst (red) configurations from Table~\ref{table:parallel}, percentage $\Delta$WCT (gain or loss) when tuning the MF value in the self-clustering heuristic \#1.}
\label{fig_perf_parallel}
\end{figure}

After characterizing the general behavior of GAIA in the parallel setup, the worst and best configurations are further investigated. For each of them, the gain/loss for the whole range of MFs is reported in Figure~\ref{fig_perf_parallel}. A MF value of $1.1$ means a large number of migrations, on the other side with MF=$19$ there are no migrations (in this specific simulation model). In terms of cost, this means that in Equation~\ref{MigrationCost} the $MigCPU$ and $MigComm$ count as zero. Therefore, this is an indirect evaluation of the execution cost of the heuristic ($Heu$). In green is reported the behavior of the best configuration and in red the worst. It is worth noting that the two configurations have two different $\pi$ and therefore are fully equivalent in terms of simulated model. The best configuration obtains a gain for all MF values up to $10$. For larger values, there is always a loss. A similar behavior can be found for the worst configuration. In this case, the tipping point is MF=$6$. In both cases, the overhead introduced by the computation of the heuristic is quite limited (less than $1$\% with $\pi=0.2$ and $2$\% with $\pi=0.5$). The monotonic behavior of both histograms (when there is a gain) will ease the development of self-tuning adaptive partitioning mechanisms.

The results for the distributed setup are reported in Table~\ref{table:distributed}. The tested configurations are the same of the parallel setup in terms of Migration Size, Interaction Size and $\pi$. In this case, there are 5 configurations in which GAIA obtains a loss. Where there is a gain, it can be quite large. As expected, the best result ($65.99$\%) is obtained when the Interaction Size ($1024$) is large and the Migration Size is at its low ($32$). In all cases in which there is a gain, the used MF is low (e.g.~[$1.1$, $2$]). Finally, the negative results have a loss that is quite limited (e.g.~$-0.38$\%) and obtained when GAIA is active but the MF is so high that there are no migrations (e.g.~MF=$19$) or a very limited number of migrations (e.g.~MF={$17$, $18$}). As in the parallel setup, the loss that is obtained when there are no migrations is an indirect evaluation of the heuristic execution cost ($Heu$).

\begin{table}[]
\centering
\begin{tabular}{ccc|cc|cc|}
\cline{4-7}
                           &                                     &                  & \multicolumn{2}{c|}{$\pi=0.2$}         & \multicolumn{2}{c|}{$\pi=0.5$}         \\ \hline
\multicolumn{1}{|l|}{GAIA} & \multicolumn{1}{l|}{Migr.~Size} & Inter.~Size & \multicolumn{1}{l|}{WCT (MF)} & $\Delta$WCT & \multicolumn{1}{l|}{WCT (MF)} & $\Delta$WCT \\ \hline
\multicolumn{1}{|c}{OFF} & \multicolumn{1}{c}{-} & \multicolumn{1}{c}{1} & \multicolumn{1}{|c}{741.00} & \multicolumn{1}{c|}{-} &\multicolumn{1}{|c}{1797.05} & \multicolumn{1}{c|}{-} \\
\multicolumn{1}{|c}{ON} & \multicolumn{1}{c}{32} & \multicolumn{1}{c}{1} & \multicolumn{1}{|c}{701.11 (1.2)} & \multicolumn{1}{c|}{5.38\%} & \multicolumn{1}{|c}{1738.97 (1.2)} & \multicolumn{1}{c|}{3.23\%} \\
\multicolumn{1}{|c}{ON} & \multicolumn{1}{c}{20480} & \multicolumn{1}{c}{1} & \multicolumn{1}{|c}{743.20 (18.0)} & \multicolumn{1}{c|}{-0.30\%} & \multicolumn{1}{|c}{1802.30 (1.2)} & \multicolumn{1}{c|}{-0.29\%} \\
\multicolumn{1}{|c}{ON} & \multicolumn{1}{c}{81920} & \multicolumn{1}{c}{1} & \multicolumn{1}{|c}{\bf 743.83 (18.0)} & \multicolumn{1}{c|}{\bf -0.38\%} & \multicolumn{1}{|c}{1799.65 (17.0)} & \multicolumn{1}{c|}{-0.14\%} \\ \hline
\multicolumn{1}{|c}{OFF} & \multicolumn{1}{c}{-} & \multicolumn{1}{c}{100} & \multicolumn{1}{|c}{849.23} & \multicolumn{1}{c|}{-} &\multicolumn{1}{|c}{1944.68} & \multicolumn{1}{c|}{-} \\
\multicolumn{1}{|c}{ON} & \multicolumn{1}{c}{32} & \multicolumn{1}{c}{100} & \multicolumn{1}{|c}{707.49 (1.2)} & \multicolumn{1}{c|}{4.88\%} & \multicolumn{1}{|c}{1765.18 (2.0)} & \multicolumn{1}{c|}{1.92\%} \\
\multicolumn{1}{|c}{ON} & \multicolumn{1}{c}{20480} & \multicolumn{1}{c}{100} & \multicolumn{1}{|c}{761.32 (1.1)} & \multicolumn{1}{c|}{10.35\%} & \multicolumn{1}{|c}{1776.87 (1.1)} & \multicolumn{1}{c|}{8.63\%} \\
\multicolumn{1}{|c}{ON} & \multicolumn{1}{c}{81920} & \multicolumn{1}{c}{100} & \multicolumn{1}{|c}{852.38 (15.0)} & \multicolumn{1}{c|}{-0.37\%} & \multicolumn{1}{|c}{1825.89 (1.4)} & \multicolumn{1}{c|}{6.11\%} \\ \hline
\multicolumn{1}{|c}{OFF} & \multicolumn{1}{c}{-} & \multicolumn{1}{c}{1024} & \multicolumn{1}{|c}{2698.50} & \multicolumn{1}{c|}{-} &\multicolumn{1}{|c}{6311.84} & \multicolumn{1}{c|}{-} \\
\multicolumn{1}{|c}{ON} & \multicolumn{1}{c}{32} & \multicolumn{1}{c}{1024} & \multicolumn{1}{|c}{998.63 (1.1)} & \multicolumn{1}{c|}{62.99\%} & \multicolumn{1}{|c}{\bf 2146.91 (1.1)} & \multicolumn{1}{c|}{\bf 65.99\%} \\
\multicolumn{1}{|c}{ON} & \multicolumn{1}{c}{20480} & \multicolumn{1}{c}{1024} & \multicolumn{1}{|c}{1065.27 (1.2)} & \multicolumn{1}{c|}{60.52\%} & \multicolumn{1}{|c}{2210.73 (1.1)} & \multicolumn{1}{c|}{64.97\%} \\
\multicolumn{1}{|c}{ON} & \multicolumn{1}{c}{81920} & \multicolumn{1}{c}{1024} & \multicolumn{1}{|c}{1272.56 (1.2)} & \multicolumn{1}{c|}{52.84\%} & \multicolumn{1}{|c}{2425.32 (1.1)} & \multicolumn{1}{c|}{61.58\%} \\ \hline
\end{tabular}
\caption{Distributed setup with self-clustering mechanism OFF/ON. Different migration and interaction sizes. Different probabilities ($\pi$) to generate new interactions. The results reported are: the mean WCT (in seconds), the Migration Factor (MF) used to tune GAIA and the gain/loss obtained by self-clustering (percentage $\Delta$WCT). In bold, the worst and best performance obtained in the distributed setup.}
\label{table:distributed}
\end{table}

\begin{figure}[h]
\centering
\includegraphics[width=8.5cm]{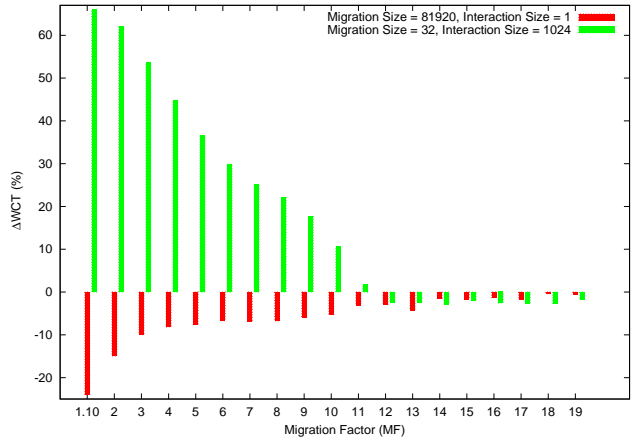}
\caption{Distributed setup: best (green) and worst (red) configurations from Table~\ref{table:distributed}, percentage $\Delta$WCT (gain or loss) when tuning the MF value in the self-clustering heuristic \#1.}
\label{fig_perf_distributed}
\end{figure}

The worst (in red) and best (in green) configurations in the distributed setup are analyzed in Figure~\ref{fig_perf_distributed}. As in the parallel setup, the best configuration has a gain for all MF$\le10$. Higher values get a loss that is near to $-3$\%. The worst configuration (e.g.~large Migration Size and small Interaction Size) for all MFs gets a loss. The loss is larger for low MF values (e.g.~$-23.91$\% with MF=$1.1$) and steadily decreases for higher MF values. Again, both the histograms show a monotonic behavior.

The absolute WCT of parallel and distributed setups are not directly comparable due to the big difference in terms of hardware specifications. Nevertheless, the results follow the expectations. The latency in the distributed architecture has a clear impact on the WCT.

To summarize, even if the tested simulation model is dynamic and with a limited amount of space-time locality, in all the tested configurations of the parallel setup GAIA is able to obtain a gain. The magnitude of this gain is limited but relevant. Clearly, better results would be obtained with less dynamic and more realistic mobility models. In the distributed setup, some specific configurations get good gains but in other cases GAIA is useless and introduces some overhead. The distributed execution is where the second part of GAIA, the computation load-balancing (or better, the combination of communication and computation load-balancing), will be fundamental for a generalized execution speedup.

\subsection{Self-tuning Adaptive Partitioning}
\label{SelfTuningAdaptPartitioning}
Up to now, the coarse tuning of the self-clustering heuristic is done exploring the space of values of a given parameter (i.e.~MF) and the fine tuning of the heuristic (e.g.~parameters such as $\kappa$ and MT, see Section~\ref{Heuristic_1}) is not considered. The next step is a self-tuning mechanism for the heuristic and, for some specific cases, an automatic selection mechanism for clustering heuristics. As a first task, we are working on the MF self-tuning. There are two main approaches that can be followed: ``inter-run'' and ``intra-run'' self-tuning. For statistical correctness of the simulation results, it is always necessary to perform multiple independent runs. Therefore, it is possible to analyze the outcomes of the previous runs to explore different configurations of the tuning parameters. By definition, each set of multiple independent runs must have exactly the same simulation model configuration and the only difference is the initial setup of the random number generator. For this reason, each run shows a different evolution of the model but it is possible to expect a high level of stability in the behavior of runs. This kind of stability, such as the monotonic behavior observed in the evaluation of GAIA, helps in the development of an ``inter-run'' self-tuning mechanisms. For the ``intra-run'' self-tuning, our approach is to analyze the behavior of the simulator during a time interval (e.g.~a certain number of timesteps) to get useful information to tune the clustering heuristic and to repeat this mechanism up to the end of the run.

\section{Related Work}\label{sec:relatedwork}

In~\cite{gda-pads-2003}, we proposed a very preliminary version of partitioning based on self-clustering. In the years, our approach has been refined and extended~\cite{gda-simpat-2014} but an accurate cost analysis has never been published.

\subsection{Partitioning of Simulation Models}
The partitioning of simulation models has been largely investigated in the past. Good introductions to the problem are given in~\cite{685264} and in~\cite{bagrodia98}. In the latter, the authors evaluate different partitioning schemes of the simulated region in order to assign an approximately equal number of simulated network nodes to each partition and an equal number of partitions to each processor, reducing the amount of inter-processor messages. In this case, the partitioning schemes are static and cannot be adjusted at runtime and furthermore static network topology is assumed. Another static approach for parallel conservative simulations is presented in~\cite{Boukerche:1994:SPM:182478.182586}. The authors of~\cite{847157} propose a combination of static partitioning and dynamic load balancing strategies. In this case, the approach relies on a conservative synchronization algorithm and can be applied only on shared memory systems.

Many of the approaches that have been proposed to deal with load balancing are designed to address computational load balancing~\cite{Alkharboush:2013:LPH:2570454.2570889} or the communication aspects~\cite{1240666,5356113} but not both of them. Moreover, very often, only optimistic synchronization (e.g.~the Time-Warp algorithm) is considered~\cite{4262808,5356113,Kim1998433}. In many of the proposed approaches, the granularity of the load balancing mechanism is at the level of the whole LP. In other words, a whole LP is migrated from a CPU to another one. A different approach is proposed in~\cite{Schlagenhaft:1995:DLB:214282.214337} in which the LP is the container of many ``basic elements'', such as happens in our approach. Also in this case, the synchronization relies on Time-Warp and even if it considers the communication cost in the initial partitioning (Corolla approach), the dynamic load balancing is based only on the CPU load.

\subsection{Partitioning of Agent-based Simulations}
An agent-based approach in proposed in~\cite{910853}. In this case, each simulated agent is implemented as a single LP. Given that, as it is difficult or impossible to determine an appropriate simulation topology a priori, a dynamic approach is implemented. The discussed solution is based on the ``spheres of influence'' (of each agent), that are used to derive an idealized decomposition of the shared state into LPs. The partitioning of the shared state is performed dynamically during the simulation. This approach supports both conservative and optimistic synchronization and can be adapted to balance both computation and communication. Apparently there are many points in common with the approach proposed in this paper, but there are also significant differences. For example, in~\cite{910853} the key parameter for rearranging the decomposition is the cost of accessing a variable for a given LP. This cost is measured as the number of other LPs that have to be traversed and the rearrangement is based on a tree-like structure of the LPs. 

\subsection{Partitioning of Multi-agent Systems}
The partitioning of Multi-Agent Systems (MAS) is a widely studied topic. The adaptive partitioning in PADS, when the simulation model is agent-based, can be seen as a specific instance of partitioning in MAS. In~\cite{Ogston:2003:MDC:860575.860702}, the authors propose a decentralized approach for the clustering of agents in large scale MAS. In this case, the partitioning is presented as a search problem in which the agents interact in a peer-to-peer fashion looking for other similar agents. In this way, the agents are partitioned in coalitions. The similarity of agents can be determined using many metrics, in this case each agent represents a two-dimensional point and the Euclidean distance is used in determining the coalitions. Clustering in autonomous multi-agent systems in which the agents are placed in a regular lattice graph is discussed in~\cite{6483388}. The agents are tagged with two colors and engage in local location-swaps. The effectiveness of a few simple strategies is evaluated using simulation. The main difference of this approach with respect to the self-clustering that we propose is the location swap of the agents. In fact, in our approach the swap is only ``logical'' (i.e.~being part of a group) and not ``physical''. Despite of this, the proposed strategies are sound and could be compared with the ones now implemented in GAIA. A framework for Multi-Agent Based Clustering is presented in~\cite{Chaimontree2012}. In this case, the agents can improve their initial cluster configuration using a two phase mechanism. A first phase based on bids and a second one with negotiations between couple of agents. In the performance evaluation, two clustering algorithms based on K-means and K-NN respectively, are considered. The approach discussed in~\cite{5299196} is relevant to GAIA. In fact, the goal of the proposed approach is to find equally sized clusters with maximal intracluster communication among agents in order to efficiently distribute agents across multiple execution unit. This is obtained analyzing the interactions among the agents and by means of online or offline processing. Another approach, strongly based on the interactions among agents is presented in~\cite{Kubera2011}. As in GAIA, in this case no distinction is made between agents and objects: all entities of the simulation are agents. In~\cite{Brooks2003} is introduced the concept of ``congregation'', that is a location and a set of agents that are all gathered in that location. The goal of creating congregations is to to efficiently find the most appropriate agents with which to interact. The aggregation process happens using a set of (physical or logical) loci that are defined at the beginning of the system lifetime. With respect to the clustering implemented in GAIA, the concept of congregation assumes a system in which the interaction pattern among agents changes less frequently.

The simulation of the movement of a large number of entities or characters (i.e.~crowds) is often designed and implemented using an agent-based approach. For example, in~\cite{Vigueras2010225}, the partitioning of agent-based crowd simulations is discussed. In this case, the agents are partitioned in irregular shape regions (convex hulls) to solve the partitioning problem. The goal of this approach is to minimize the number of agents near the borders of the regions, and to properly balance the number of agents in each region too. Even if there are some similarities with the approach that we propose, there is a big difference in the partitioning method. That is, in our case the simulated space is not partitioned in regions and the shape of clusters is free and it is determined only by the interaction of the simulated entities at runtime. The crowds partitioning problem is also discussed in~\cite{5429649}, in this case an adapted k-means clustering algorithm is used for the partitioning. More specifically, the partitioning is based on the position of agents in the simulated world. The k-means algorithm is executed at predefined intervals to check if the partitioning quality is acceptable or if a new partitioning is necessary. In~\cite{Al-Zinati:2015:SVE:2772879.2773283} a virtual environment for agent-based simulations is partitioned into areas called cells. The virtual agents are situated in the virtual environment while specialize agents have the role of controllers and coordinators. The specialized agents re-organize them-selves and the environment structure to ensure that the simulation functional and performance requirements are met. Just like in the self-clustering approach presented in this paper, the virtual agents are unaware of the partitioned structure of the environment and the underlying self-organization activities. A big difference with respect to our proposal is the presence of cell boundaries that, in case of hotspots, require the splitting of cells. 

\subsection{Computational and Communication Aware Partitioning}
A relevant work is presented in~\cite{4262808}, in which the authors propose a dynamic partitioning algorithm which performs both computation and communication load balancing. In this case, the simulation model is composed of modules, to each of which is manually assigned a weight, and the modules are grouped into LPs. The initial partitioning has the goal to balance the computation load according to the resources that are available on each host that is participating to the simulation. To estimate the capacity of each host, a performance benchmark is executed before starting the simulation runs. After this preliminary phase, the final assignment of LPs to hosts is done using a heuristic bin packing algorithm. The dynamic part of the approach is based on measurements at runtime and the reallocation (i.e.~migration) acts on whole LPs. This is done in a sequence of dedicated cycles such as the ``communication refinement cycle'' that is followed by a ``load balancing cycle''. Also in the case of this approach, an optimistic synchronization algorithm is assumed. Another approach that takes into account both computation and communication is presented in~\cite{1348301}. In this case, benchmark experiments are used to take into account the specific distributed computation environment and the specific details of the network model (e.g. network topology and traffic characteristics). The benchmarks are executed before the simulation runs and there is no adaptation at runtime, in other words the partitioning is static. In~\cite{Nandy:1993:PPT:158459.158465} is assumed that there is complete knowledge of computation and communication load before the simulation is executed. This is very different from our approach, in which we do not assume a priori knowledge of both the execution architecture and the simulation model behavior. A very specific approach is used in~\cite{Boukerche:2000:PPS:510378.510591}, that employs a two-stage parallel simulation that makes use of a conservative scheme at stage 1 and of Time-Warp at stage 2. The goal of the first stage is to collect data that will be used to partition the simulation and improve the balancing in stage 2. The improvement of the second stage is obtained stabilizing the Time-Warp and therefore reducing the rollback overhead. In~\cite{Boukerche20021454}, the authors propose a simulated annealing algorithm to find good (sub-optimal) partitions for execution on a multi-computer. In this work, the synchronization is demanded to the Chandy-Misra null-message algorithm and the objective function of the partitioning is chosen so that inter-processor communication conflicts are minimized, processor load remains balanced, and the probability of sending a null message between processors is minimized. It is worth noting that, in this approach, the basic unit of the dynamic balancing is the LP (considered just like a particle moving in a physical space) and the balancing is done assessing the ``force'' of interaction between the LPs. The mechanism evaluated in~\cite{745025} has some similarities with our proposal. In fact, the proposed dynamic partitioning is based on object migration and takes account of both communication and computation. Despite this, there are also relevant differences. First of all, the method in~\cite{745025} is based on set of assumptions (i.e~the number of hosts running the distributed simulation is in the range from 4 to 8; the fluctuation of the network bandwidth is negligible; the migration cost of a piece of workload is proportional to the physical size of the migrating workload). Furthermore, in this case the process of choosing what objects must be migrated is partially centralized. On the other hand, the heuristic clustering mechanism that we propose is not based on such kind of basic assumption and its architecture is fully distributed.

\section{Conclusions}\label{sec:conclusions}

In this paper, we have studied the model partitioning problem in parallel and distributed simulation. An adaptive solution that is based on the self-clustering of simulated entities has been proposed. The support for adaptive parallel and distributed simulation has been implemented in a simulator called GAIA/ART\`IS~\cite{pads}. A specific cost analysis for parallel and distributed simulations has been introduced and used for the performance evaluation of the proposed middleware (and adaptive approach). The results, in both parallel and distributed execution architectures, demonstrate that, under the communication viewpoint, GAIA is able to obtain good results even with a simulation model that is not favorable.

\section*{Acknowledgments} 

We would like to thank the anonymous reviewers whose detailed comments (and suggestions) greatly contributed to improve the overall quality of this paper.

\appendix
\section{Table of Symbols or Acronyms}\label{app:symacro}
The main symbols and acronyms used in the paper are reported in Table~\ref{tab:symbols}.

\begin{table}[h]
\begin{center}
	\begin{tabular}{ | l | p{8cm} |}
	\hline
	\textbf{Symbol or Acronym} & \textbf{Description} \\ \hline
	$\alpha$ & Ratio of \emph{internal} vs. \emph{external} interactions \\ \hline
	CC & Communication Cost \\ \hline
	$\epsilon$ & Amount of \emph{external} interactions \\ \hline
	Heu & Cost of the heuristic function \\ \hline
	$\iota$ & Amount of \emph{internal} interactions \\ \hline
	$\kappa$ & Heuristic \#1. Observation time \\ \hline
	LCC & Local Communication Cost \\ \hline
	LCR & Local Communication Ratio \\ \hline
	LP & Logical Process \\ \hline
	MCC & Model Computation Cost \\ \hline
	MF & Migration Factor \\ \hline
	MIC & Model Interaction Cost \\ \hline
	MigC & Migration Cost \\ \hline
	MigComm & Cost to transfer the migration messages \\ \hline
	MigCPU & Generic migration computation time \\ \hline
	MMC & Middleware Management Cost \\ \hline
	MR & Migration Ratio \\ \hline
	MT & Migration Threshold \\ \hline
	$\omega$ & Heuristic \#2. Number of interactions considered \\ \hline
	$\pi$ & Probability  that, in a given timestep, a SE generates a new interaction \\ \hline
	PEU & Physical Execution Unit \\ \hline
	RCC & Remote Communication Cost \\ \hline
	SE & Simulated Entity \\ \hline
	TEC & Total Execution Cost \\ \hline
	SC & Synchronization Cost \\ \hline
	$\zeta$ & Heuristic \#3. Interactions since the last evaluation \\ \hline
	WCT & Wall Clock Time \\ \hline
	\end{tabular}
\end{center}
\caption{Table of Symbols and Definitions}
\label{tab:symbols}
\end{table}

\bibliographystyle{elsarticle-num}
\bibliography{paper}

\end{document}